\magnification=1200
\baselineskip=17.0truept
\noindent \overfullrule=0pt \par
\bigskip
\centerline{{\bf On The Distribution of Zeros }}
\centerline{{\bf of Chaotic Wavefunctions}}
\bigskip
\centerline{{\bf Pragya Shukla}}
\centerline{{Condensed Matter Theory Unit, Jawaharlal Nehru Centre 
for Advanced Scientific Research,}}
\centerline{{Indian Institute of Science, Banglore-560012, India}}
\bigskip
\bigskip
\bigskip
\bigskip
\centerline{{\bf Abstract}}
.
The wavefunctions in phase-space representation can be expressed
as  entire functions of their zeros if the phase space is compact. 
	These zeros  seem to hide a lot of relevant 
and explicit information about the underlying classical
dynamics. Besides, an understanding of their statistical properties 
may prove useful in the analytical calculations of the wavefunctions in 
quantum chaotic systems. This motivates us to persue the 
present study which by a numerical statistical analysis shows that both 
long as well as the short range correlations exist between
zeros; while the latter turn out to be universal and
parametric-independent, the former seem to be system dependent
and are significantly affected by various parameters i.e
symmetry, localization etc.
Furthermore, for the delocalized quantum dynamics, the distribution of 
these zeros seem to mimick that of the zeros of the random
functions as well as random polynomials.\par


\vfill\eject
\par
\vskip 5mm
\noindent{{\bf 1.Introduction}}
\vskip 5mm
	In this paper, our aim is to study the statistical distribution of
	zeros of "quantum chaotic" wavefunctions, expressed in
	phase-space representation. 
	The wavefunctions of quantum chaotic systems
	have so far remained a relatively less explored (as compared to
	spectra) area notwithstanding their importance in quantum
	mechanical as well as semiclassical analysis. 
	The
	few of their known properties are 
	the microcanonical nature of wavefunctions if
	the underlying classical dynamics is chaotic [1], the scarring 
	behaviour [2,3], dynamical
	localization [4] or a few statistical (numerical) studies of the
	components of these wavefunctions [5,6].
	The relevant information hidden in the semiclassical
	wavefunction about the underlying quantum-classical
	correspondence is not yet fully extracted. This motivates us to
	pursue the present problem. As intution suggests us,
	this can be best studied by analyzing various properties of the 
	wavefunctions in the phase-space where 
	quantum dynamics in limit $\hbar\rightarrow 0$ can be compared with
	the underlying classical dynamics. One such analysis involves the 
	study of the distribution of zeros (or nodal patterns) of 
	wavefunctions (in phase space representation).\par 
		The interest in the phase space study of the nodal patterns
	arises following a study by Leboeuf and
	Voros [7,8] where they show that the coherent state representation of
	a wavefunction (i.e Bargmann or
	Husimi function)
	has a finite number N of zeros
	in phase space (N being the dimension of Hilbert space) and the
	state of the system can be completely expressed in terms of
	these zeros. Furthermore, their numerical studies on quantum maps
	(e.g Baker and kicked rotor) have also revealed that the
	zeros, in the semiclassical regime $\hbar\rightarrow 0$, condense on
	lines for classically integrable systems whereas for strongly
	chaotic quantum maps they appear to diffuse fairly
	uniformly over the phase space; this behaviour can  be explained
	semiclassically [7,8]. The only exception to this behaviour, so 
	far, has been seen for one eigenfunction in
	$N=2^k$ case of the Baker map (strongly chaotic) where zeros have a 
	dominantly linear
	distribution. In general, the distribution of 
	zeros seems to mimic the underlying classical dynamics in
	semiclassical limit (e.g. for those chaotic cases of the kicked
	rotor, where
	quantum dynamics is localized in momentum space, zeros were
observed to be localized). This makes it relevant to seek
further information about their behaviour.\par 
		In principal, we have an analytical formulation 
available to write the eigenfunction directly in nodal pattern 
representation, (a polynomial for spherical phase space and a
complicated function for torus and cylindrical phase space) [7].
But this formulation does not help us very much as it depends on the 
knowledge of the 
	position of each zero in phase space which is analytically difficult
	to evaluate. Still it can be used to
	determine the statistical behaviour of
	the eigenfunctions if the 
	statistical properties of zeros are known.\par

	The random nature of chaotic wavefunction and its  
	polynomial form in Bargmann representation (in a
spherical phase space only) has motivated people to seek analogy
between the distribution of it's zeros and roots of random
polynomials (later referred as RP) [9]; recently the statistical
properties of the latter case were analytically explored by us [10].
Our results showed an excellent agreement with the corresponidng
ones for the kicked top and the kicked rotor. It was a little
surprising in the latter case where dynamics is confined to a
torus and an arbitrary wavefuction is a complicated function
instead of a polynomial; one should expect the zeros of random
functions (superposition of functions with random weight, later
referred as RF) to be
a better model. This seemed to hint that the zeros,
while distributing themselves in semiclassical limit, do not
distinguish between spherical and torus phase space [10]. 
In other words, the zeros of RF and RP seem to have the same
distribution properties. As we show in this paper, this is
indeed the case.  We also extend the above study to understand the effect
of various parameters (i.e symmetry-breaking, localization etc.) on the 
distribution of zeros. We expect this due to our prior knowledge of the
significant influence of
these parameters on quantum dynamics [4,11,12]. \par  
   	 We choose the kicked rotor
   system for this purpose as it 
   has been an active model of research, containing a 
   variety of features such as localization, resonance, dependence of
   the spectra on number theoretical properties etc and has been used as a
   model for a very wide range of physical systems [4]. Besides, the kicked
   rotor can also display discrete symmetries such as parity and time
reversal [5,11,12]. This will enable us to study the way these 
symmetries reveal themselves in the distribution of zeros.\par
	  
	This paper is organized as follows. The section 2 contains a 
brief review of quantum properties of kicked rotor. The section 3 briefly
describes the various phase space representations and their advantage
over each other. The section 4 deals with the numerical analysis of 
distribution of zeros for kicked rotor for various parametric conditions,
 followed by a conclusion in section 5.
\vskip 5mm
\noindent{{\bf 2. The Kicked Rotor: Classical and Quantum Dynamics}}
\vskip 5mm
	   The kicked rotor can be described as a pendulum subjected to
periodic kicks (with period T) with the following Hamiltonian

$$ H = {{\left(p+\gamma\right)^2}\over 2} + {K\over 4\pi ^2}\quad {\cos}
(2\pi
\theta+\theta _0) \sum_{n=-\infty}^{\infty} \delta (t-nT). \eqno (1)$$
where $K$ is the stochasticity parameter. The parameter $\gamma$ and
$\theta_0$ are introduced in the hamiltonian in order to mimick the
effects of the time-reversal (T) and the parity (P) symmetry breaking
in the quantum hamiltonian while acting in a finite Hilbert
space.\par
        Floquet's theorem enables us to describe the related quantum 
       dynamics
by a discrete time evolution operator $U=GBG$ where $B={\rm
exp}(-i K \cos(2\pi\theta+\theta_0)/4\pi^2 \hbar)$ and 
$G={\rm exp}(-i(p+\gamma)^2/4\hbar)$. Here $\theta$ and $p$ are
position and momentum operator respectively, with $p$ having discrete
eigenvalues $p|l>=l\hbar|l>$ due to periodicity of $\theta$, 
($\theta=\theta + 2\pi$). The nature of the quantum dynamics and
therefore the statistical properties of the
associated quantum operators depend on $\hbar$ and $K$. The choice of
a rational value for $\hbar T/2\pi$ results in a periodicity also for
the momentum space ($p=p+2\pi M$ or $l=l+N$) and, therefore, in
discrete eigenvalues for $\theta$ ($\theta |n>={2\pi n\over M}|n>$).
The quantum dynamics can then be confined to a 2-dimensional torus 
(with finite Hilbert space of dimension $N$) and $U$ can be reduced
to a finite $N$-dimensional matrix of the form [7,4]

$$ \eqalignno{U_{mn} & = {1\over N^2} \sum_{j}{\rm exp}{\left[-i{K
\over 4\pi^2 \hbar}cos{\left({2\pi j\over
N}+\theta_0 \right)}\right]} & \cr
& \sum_{l,l'=-N_1}^{N_1} 
{\rm exp}
\left[-i\left({\pi ^2 \hbar}(l^2+l'^2) 
- \pi \gamma (l+l')
\right)\right] 
 {\rm exp}\left[-i\left( {{l(m-j)+l'(j-n)}\over N}\right)\right]
,& (3) \cr }$$
where $N_1 = (N-1)/2$ (with $N$ odd) and $m,n = -N_1, -N_1 +1,...,N_1$ .\par
	In contrast to the classical dynamics, the quantum dynamics can be
	significantly affected by 
the relative values of three parameters, namely, $K, \hbar$ and $N$.
It has already been pointed out [4] that, for $K^2 > N\hbar^2 (=2\pi
M/T)$, the eigenstates are fully extended in momentum space (later
referred as delocalized case); it is also termed as strong
chaos limit as here the kick is strong enough to make the
classical motion ergodic in whole of the phase space [12]).
We also know that the spectrum as well as the distribution of
eigenvector-components in this case,
with parameters $\gamma$ and $\theta_0$ chosen to preserve
either exact or partially violated symmetry, can be modeled by the random
matrix theories (RMT) [5,6,11,12]. Under this limit, the quantum dynamics has
 a time
reversal symmetry $T$ for $\gamma=0$ and a parity symmetry $P$ for $\theta_0
=0$. Though the $T$-symmetry may be violated for $\gamma\neq 0$, still a more
generalized antiunitary symmetry $S=TP=PT$ can be preserved in the system if
$\theta_0 =0$ [5,11]. By a slow variation of these parameters, one can realize
the various intermediate stages of the statistical properties of quantum
operators, for example, the Poisson $\rightarrow$ COE transition of the
spectrum for K-variation, COE $\rightarrow$ CUE for $\gamma$-variation
with $\theta_0$ fixed at nonzero value etc.
In the opposite limit of weak chaos, 
namely, $K^2 << N\hbar^2$ (corresponding to a
diffusively covered phase space with a non-uniform
distribution of periodic orbits), the
eigenstates localize in the momentum space and one obtains a
Poisson distribution for the spectrum (later referred as
localized case). Moreover, here the symmetry
breaking may not be realized in the quantum dynamics just by making
$\theta_0$ or $\gamma$ sufficiently non-zero.\par
\vskip 5mm
\noindent{{\bf 3. The Phase Space Representations of Wave Function}}
\vskip 5mm
	The phase space representation of the eigenfunctions of
quantum mechanical operators can give excellent information about the
semiclassical description of a quantum state. A semiclassical wavefunction 
can be constructed, first by writing the quantum state as a function of the 
parameter $\hbar$, and then by exploiting the fact that this function 
should approach the corresponding classical function in the limit 
$\hbar\rightarrow 0$ [13].
	This is easy
	to be observed in the phase space representation of
	quantum state, where the quantum dynamics (the Heisenberg equation),
	analyzed in terms of density operator $\rho = |\psi><\psi|$,
	explicitly appears as a deformation of classical dynamics (the
	Liouville equation). This enables one to compare the quantum
	function with the corresponding classical function, order by
	order in $\hbar$. The inversion $W \rightarrow \psi $ then
	gives, in principle, the quantum state for various orders in
	$\hbar$.
	The two most widely used phase space techniques are the Wigner
	representation $W_w$ defined as

$$ W_w (q,p,\hbar) = (2\pi\hbar)^{-N} \int{\psi (q-{r\over 2})
   \psi(q+{r\over 2}) {\rm e}^{ipr/ \hbar} {\rm d}^N r}, \eqno (4)$$
and the Husimi representation $W_h$ given as

$$ W_h (q,p,\hbar) = {{|<z|\psi>|^2} \over {<z|z>}}, \eqno (5)$$
where $|z>$ is a plane coherent state described by following standard
formulas,

$$ |z> = {\rm e}^{{\bar z}a^+} |0>,\qquad a^+ = {(\hat q - i \hat p)\over
 {\sqrt 2}\hbar},\qquad z = {q-ip\over \sqrt 2}, \eqno (6) $$
with $q$ and $p$ as the position and momentum eigenvalues of the operator
$\hat q$ and $\hat p$ respectively and its overlap with position wave 
function given as follows,

$$ \langle {q|z}\rangle = (\pi \hbar)^{-N/4} {\rm exp}
{\left[\left(-({\hat z}^2 + q^2) + \sqrt 2 \hat z q \right)/\hbar \right]}.
 \eqno (7) $$
Both of these representations have a nice semiclassical behaviour,
converging to the classical phase space distribution in the limit $\hbar
\rightarrow 0$. But the determination of this convergent form of $W_w$
and $W_h$ can be done only after smoothing out the oscillations
 (stronger in the case of $W_w$)
existing in both cases. In general, however (i.e for chaotic cases) even
this knowledge is not sufficient to reconstruct the wavefunction itself
by inverting the map $W
\rightarrow \psi$. The reason is that the quantum phase, which controls
the fast oscillations, is obliterated in the limiting process and can not be 
easily regenerated.
Moreover these representations also run 
into difficulties when applied to
compact phase spaces. It is because the
existence and uniqueness of the construction of Wigner representation,
which imposes itself by symmetry on a linear phase space, is no longer
clear on a compact spaces [7].  One does not face these 
difficulties with Husimi representation which only needs coherent states.
But the usual plane coherent state, used in Husimi  formulation, becomes 
nonanalytic when restricted to a compact space [7]. 
 Thus, for our purpose, we should choose a coherent state with analytical 
properties in compact space. \par
As discussed in [7], the analyticity of quantum state can be preserved
by expressing it in "Bargmann representation". This representation
diagonalizes the creation operator $\hat z$ and expresses a state as
function $\psi(z) = <z|\psi>$ with $\hat z^{+} = \hbar {{\rm d} \over
{\rm d}z} $ (similar to Schorindger representation $\psi(q)$ of quantum
state, obtained by diagonalizing the position operator $\hat q$ with
$\hat p = -i\hbar {{\rm d}\over {\rm d}q}$). The function $\psi(z)$,
describing  
the overlap of wavefunction $\psi$ with the
coherent state $|z>$, belongs to a Hilbert space of entire (analytic)
functions with finite norm given as follows,

$$<\psi(z)|\psi(z)> = (2\pi \hbar)^{-1} {\int |\psi(z)|^2 {\rm
e}^{-|z|^2 /\hbar} {\rm d}q {\rm d}p}. \eqno (8) $$
Furthermore the Husimi function can be expressed in terms of $\psi(z)$
as $W_H = |<z|\psi>|^2 / <z|z>$. Thus one can consider $\psi(z)$ as a
sort of phase space representation for the wave vector $|\psi>$.
The advantage of
working with $\psi (z)$ is that it preserves the information about the 
quantum phase explicitly.\par
    It has been shown in ref.[7] that the antianalytic coherent
    states for compact space such as the torus (the case considered
    here) can be obtained by applying  periodic boundary conditions on
    the
antianalytic coherent states $|z>$ on the plane. 
 Thus the imposition of the appropriate periodic boundary conditions 
 on the kernel (5), depending on the
 nature of the compact space, leads to a Hilbert space of entire
 functions $\psi(z)$. 
 For our
 purpose, we only mention here the case of the torus,
 where application of the periodic boundary conditions in  both the
 $q$ as well as $p$ directions  and  use of the condition
 $2\pi N\hbar = 1$ results in following form for $\psi(z)$

 $$ \psi(z) = \sum_{n=0}^{N-1} \psi(q_n) <<z|q_n>> \eqno (9) $$
 where $\psi _n$'s are the eigenfunctions of operator $U$ in the coordinate
 representation and $<<z|q_n>>$ is projection of the coordinate
 eigenfunction $|q_n>$ in the $z$-space, under the periodic boundary
 conditions.

 $$ <<z|q_n>> = (2N)^{1/4} {\rm exp}\left[-2\pi N \left((z^2 + q_{n}^2)/2
 - {\sqrt 2}z q_n \right)\right] 
 \theta _3 \left(i\pi N \left[ q_n - i{\theta _q \over 2\pi N} - 
 z \sqrt 2 \right] | iN \right) \eqno (10) $$
 Here coefficients $\psi(q_n)$'s are complex under no-symmetry condition.
But presence of a T-symmetry makes them real as now if $\psi(z_p)=0$ then 
$\psi(Tz_p)$=0 too [10]. The existence of an additional symmetry such as 
parity 
imposes further restrictions on $\psi(q_n)$'s; $\psi(q_n)=0$ if $n$ is odd 
and $\psi(q_n)\not=0$ if $n$ is even.
 Further, by using the periodic boundary conditions
 leading to this compact space, it can be shown [7] that every function
 $\psi(z)$ has exactly $N$ zeros in the fundamental domain (defined by 
 $\delta q=1, \delta p=1$).\par
It is worth reminding ourselves here that, for a spherical phase
space, $<<z|q_n>>$ takes the form of a polynomial [7] and 
therefore the choice of $\psi(q_n)$'s as Gaussian random variables leaves 
$\psi(z)$ as a random polynomial. In ref.[10], we had analytically
calculated the joint probability density of these zeros as well as the
correlation function and compared the result with numerical results for 
kicked top (which has a spherical phase space); we found a good
agreement. But as obvious from eq.(10), for a 
torus phase space, $\psi(z)$ is not a polynomial. A choice of
$\psi(q_n)$'s as random variable gives rise to a random 
function (later referred as RFC for complex and random coefficients,
RFR for real and random coefficients and RFA if the coefficients
are real, random
and non-zero for even $n$ and zero for odd $n$).  

\vskip 5mm
\noindent{{\bf 4.Numerical Analysis}}
\vskip 5mm

           With this knowledge, we can
	now proceed in following steps, 
	(i) numerical computation of
eigenfunctions by diagonalizing the matrix U (eq.(3)), (ii)
calculation of their Bargmann transforms, using eq.(9),
(iii) numerical computation of the position of
each zero using Cauchy's integral (see [7]), 
(iv) the, so obtained, $N^2$ zeros ($N$ for each of the $N$
eigenfunctions) are then used for
evaluation of the various statistical measures. For the reasons 
explained above, We repeat the steps (iii) and (iv) also for
random functions cases,
with normally distributed coefficients $\psi(q_n)$'s.\par
	To see the effect of various parameters on the
statistical behaviour of zeros, the above steps are repeated for
many of their combinations. Here, untill otherwise stated, we always 
take  $(K=20000, M=1, N=199,
\gamma =0.7071, \theta _0 = \pi/2N)$ for delocalized case with no-symetry, 
$(K=20000, M=1, N=199,\gamma =0.0, \theta _0 = \pi/2N)$ for
delocalized case with one symmetry (time-reversal), $(K=20000, M=1, N=199,
\gamma =0.0, \theta _0 = 0.0)$ for delocalized case with two
symmetries (namely time-reversal and parity) and  $(K=10, M=30, N=199,
\gamma =0.7071, \theta _0 = \pi/2N)$ for localized case with no symmetry.
To improve the statistics, we
superimpose the statistics for $N$ zeros of each of the $N$ eigenfunctions
and set T=1.\par
	We start by computing the density $\rho$ of zeros, 
distributed over
the phase plane of unit area. We know that the eigenfunctions of RM 
ensembles, in Bargmann representation, have the form of the random
polynomials. The nature of the coefficients in these polynomials
depends on the symmetries of the ensembles (e.g complex for no
anti-unitary symmetry and real for 1
anti-unitary symmetry).
Recently it was shown that
the zeros for RP with complex coefficients are uniformly
distributed in the phase space for any arbitrary N while, in 
case with real coefficients, this uniformity no longer survives
due to an increased concentration of zeros on real axis [10].
The same should hold good for the eigenfunctions of
quantum systems with strongly chaotic classical dynamics where a
typical wavefunction randomly goes all over the  phase space
and has a uniform intensity distribution except for a few scars (high 
intensity
regions) [1,2]. This expectation has its roots in the success of
RMT in modelling 
"quantum chaotic" spectra; we hope the same for "quantum chaotic"
eigenfunctions too. 
In fact, the figure $1(a)$ which shows
the distribution of $99\times 99$ zeros of all $99$ eigenfunctions of QKR
$(K=20000, M=1, N=99,
\gamma =0.7071, \theta _0 = \pi/2N)$
strongly suggest the uniformity of density
in the phase plane. The
various parameters here are chosen so as to ensure the underlying
classical dynamics to be strongly chaotic and removal of all symmetries
from the quantum case [5,11,12]. The figure $1(b)$ shows the case with a 
chaotic classical dynamics while the quantum dynamics is localized
in the momentum space 
$(K=10, M=30, N=99,
\gamma =0.7071, \theta _0 = \pi/2N)$. 
 Here again the zeros seem to be randomly and
uniformly distributed, except for an increased concentration on 
parity symmetry axis $q=0.5$ (reasons explained later). The
apparent randomness and uniformity of the
distribution of zeros (away from the axis) in this case is
indeed misleading and occurs due to the
averaging effects produced by superposition of many wavefunctions. 
In fact, the figure $2(a)$, where
the Husimi distribution of a single eigenfunction is plotted, shows that
the zeros for the localized case, in general, form patterns; for few of the
eigenfunctions, zeros seem to be localized too. It seems that
the maximum intensity regions, in order to concentrate in some part of
phase space, compel minimum intensity regions to distribute in an
ordered way. 
This pattern formation
of zeros is more obvious for the eigenfunctions of the localized dynamics
with a symmetry (figures $2(b),2(c)$). The figure $2(d)$ shows the
Husimi plot of an eigenfunction for delocalized case. Here zeros of even
a single eigenfunction seem to be uniformly distributed. In fact, 
the distribution approaches the uniformity
in $N\rightarrow\infty$ i.e the semiclassical limit (unlike the RMT case
where uniformity exists for all N) and therefore
mimicks the behaviour of underlying strongly chaotic classical
dynamics. 
\par
In order to see whether the density $\rho(p,q)$ is separable in
variables $p$ and $q$ or not, we calculate the coefficient of
correlation $C(m,n)$, defined as follows,

$$C(m,n) = {{<p^m q^n> - <p^m><q^n>}\over {\sigma (p^m) \sigma (q^n)}},
 \eqno (11) $$
 where $\sigma^2 (r^m)$ is the variance of $m$th moment of the 
variable $r$. To see the effect of various parameters as well as
dimensionality, we calculate $C(m,n)$, $(m+n<8)$, for
delocalized case with no symmetry, one symmetry and two symmetry
respectively and also for localized case with no symmetry, with each
 case studied for four values of $N$, namely, $N=61,
99, 149, 199$. We find that, for all the parametric cases, the
correlations between $p$ and $q$ variable decrease with $N$. In fact,
the correlations for lower moments are much less than one sample
error even for $N=61$ (with $C(m,n) \leq 10^{-5}$ for $m+n \leq
5$ and sample error =$10^{-2}$), thus indicating their
near-absence even for  small $N$-values. The correlations for
higher moments $(m+n>5)$ are although not negligible for small
$N$-values but decrease rapidly with increasing $N$ and can 
safely be assumed to be zero in limit $N\rightarrow\infty$, (e.g 
$C(4,4)=0.018\pm0.016$ for $N=61$ and $0.003\pm0.005$ for
$N=199$, for delocalized case without symmetry). 
This implies that, for the delocalized as well as the localized case,  
we can study the distribution of zeros in either $p$ or $q$
direction without taking care of $q$ or $p$ variable, respectively.
Thus, for a check on uniformity, it is sufficient to study the measure 
$<\rho>_q$
and $<\rho>_p$ (where $< >_r$ implies the average over variable $r$). 
As shown in figures $3(a)$ and $3(b)$, the averaged density  in
both $q$ (averaged over $p$) and $p$ directions (averaged over
$q$) for delocalized case is nearly a
constant, thus indicating a uniform distribution while, for 
the localized case, it oscillates strongly about an average value. 
Here the increased average density around some values of $p$ and $q$ 
implies the preference of zeros for this value; this behaviour may be
attributed to the localization and pattern formation of zeros.
 For the delocalized case, the
oscillation strength decreases rapidly with increase in $N$ which 
indicates the uniform density also in $p$-direction, in large-$N$ limit.
For comparison, we have also plotted the corresponding RF case in both
figures $3(a)$ and $3(b)$; the set of normally distributed $\psi(q_n)$'s is 
generated bu using "GASDEV" subroutine [14].
(note, the similarity of the delocalized case 
with the RF case in figure 3(a),(b)).
\par
 To see the effect of symmetry on the distribution of zeros, we study it
for various values of parameters $\gamma$ and $\theta _0$. 
The figures $4(a)$ and $4(b)$ show  the distribution of the
 $61\times 61$ zeros of all 61
eigenfunctions of QKR for 
$(K=10, M=30, N=61, \gamma =0.0, \theta _0 =0)$ 
and  the case $(K=10, M=30, N=61,
\gamma =0.0, \theta _0 = \pi/2N)$. Here the first set of values 
correspond to preservation of both $T$ and $P$ symmetries in the system 
whereas in the second case only $T$ symmetry is preserved.
The figure $4(a)$ shows the increased
density of zeros on the two symmetry axis (namely, $p=0.5$ and $q=0.5$)
while in figure $4(b)$ the zeros are concentrated 
only on one symmetry axis; the
concentration of zeros also on $p=1.0$  is due  to the torus
boundary conditions. Moreover zeros on either side of the symmetry
axis seem to avoid it, thus creating a hole close to
and around the symmetry axis.
This phenomenon, which occurs in the case of RP (and RF) too,
can be explained as follows:
if $\psi(z_p)=0$, then $\psi(Rz_p)=0$, where $R$ refers to a particular
symmetry in the quantum system and $Rz_p$ the $R$-symmetry counterpart
of $z_p$.  This implies that, given $z_p$ a zero of wavefunction $\psi$,
$Rz_p$ will also be its zero; zeros either come by pairs symmetric
with respect to the symmetry-axis or they are single and lie on the axis.   
Although this destroys the uniformity of distribution of zeros on a global 
level, locally and away from the symmetry axis the uniformity is still 
preserved (figures $3(c,d), 4(a,b)$).
It should also be noticed in figure $1(b)$
 that though both $\gamma$
 and $\theta_0$ 
are non-zero still the density of zeros is higher along q=0.5 line while
this behaviour is absent in delocalized case for the same values of
symmetry breaking parameters (figure $1(a)$). It seems that for
localized case, a more generalized symmetry remains preserved even
for non-zero values of $\gamma$ and $\theta_0$. 
\par
       In order to analyze the correlations between these
zeros as well as to know the extent of randomness, we compute the 
2-point correlation function $R_2(r)$.
It is defined as the joint probability-density of finding a zero at a radial
distance $x$ from the origin and another at $x+r$, averaged over all 
$x$ and $\theta$ where $\theta$ is the angular dependence of the
zero. Here $r$ is measured in units of averaged spacing and $R_2 (r)$ is 
unfolded in such a way so that average number of zeros at a distance $r$ 
 from a given zero does not depend on $r$. In Ref.[10], we had 
analytically obtained $R_2(r)$ for zeros of 
random polynomials; a numerical comparison of this formulation with 
that for kicked spin system as well as kicked rotor showed an 
excellent agreement. Here we numerically compare $R_2(r)$ for delocalized-QKR
with that for the random functions
(with complex coefficients to model the no-symmetry case and real 
coefficients for time-reversal symmetry preserving case) and find the 
latter to be a good model for the former; see figures 5(a),(b),(c). 
The agreement of 
results also confirms that zeros for both random polynomials and random 
functions, in $N\rightarrow\infty$, have the similar correlations at
least up to the second order.
The histograms in figures 5 (a),(b),(c),(d) also indicate that, for short
distances, the two-point correlation 
increases as a power law with distance, attains a maximum value and then
oscillates rapidly around this average value when distances are large.
Thus, for short distances $(r<1)$, zeros seem to repel each
other 
; the repulsion is weaker
for localized case than the delocalized case (figure 5(d)). Furthermore the 
presence of a symmetry weakens the repulsion at short distances but does 
not seem to affect the correlations at large distances (figures
5(b),(c),(d)) (but as shown later, long range correlations are indeed
symmetry-dependent). Our comparison of short-range correlations under
various symmetry-conditions for many $N$-values ($61,99,149,199$)
seems to suggest the decrasing tendency of differences in the
repulsion-strength under various parametric conditions with
increasing $N$. We believe that no difference should survive in
$N\rightarrow\infty$ limit; this is also corroborated by our
spacing-study given later. Moreover the 
short-range ($r<1$) correlation between zeros is different from that for
the two dimensional random distribution of points with $R_2(r)=1$ but 
behaves in a similar fashion for $r>1$. Note here that the long range
behaviour for all the cases approaches $1$, although oscillating
heavily around it, irrespective of the
parameteric conditions or the type of the distribution. These
oscillations are not insignificant; they should contain information
about the differences in long-range correlations under various
parametric conditions so as to agree with our number-variance study,
given later. \par

The $R_2(r)$-study suggests the short-range correlations to be free
from the influence of various parameters. 
To understand this better, we carry out a study
for the nearest neighbour spacing distribution $P(s)$ of zeros
(with all geometrical factors scaled out; see [10]) for both
QKR as well as random functions. A comparison of the two reconfirms the
conclusion obtained by $R_2(r)$-study, that is, the short-range 
fluctuation-measures of zeros-distribution of random
functions can very well model that of QKR (figures 6(a),(b),(c)). 
To understand the 
functional behaviour of $P(s)$, we fit the following equation 
$$ P(s) = \alpha s^a \quad {\rm exp}[-\beta s^b] \eqno (14)$$ 
where $a$ and $b$ are the fitting parameters and $\alpha$ and $\beta$ 
are obtained by imposing following normalization conditions, 
$\int_0^2 P(s) {\rm d}s =1$ and $\int_0^2 sP(s){\rm d}s =1$.  
The fitting analysis carried out for four values of $N=61,99,149,199$
 leads us to expect $a=1.5$ and $b=5$. Using these values as a hypothesis 
for $P(s)$, We perform the $\chi^2$-analysis which gives us 
$\chi^2=12.56$ for $N=199$ and $14$ degrees of freedom at $5\%$ level 
of significance, a value much less than the theoretical
$chi^2$-value ($=23.68$). This
therefore indicates the correctness of our hypothesis. We also find
that the calculated $\chi^2$-value decreases with increasing $N$
(e.g. $\chi^2=148.18$ for $N=99$ and $14$ degrees of freedom) which implies 
the tendency of P(s) to approach our hypothesis in $N\rightarrow\infty$
limit. \par
  Our $P(s)$-analysis also shows that the short range correlations
between the zeros are not of the same type as those for a two-dimensional
random distribution of points ( with $P(s)$ as two dimensional Poisson
distribution with $a=0$ and $b=2$ [14]). In fact, it has been
indicated [8] that the apperance of a term of the type of general
n-body interaction in the dynamics of zeros should produce a strong
short-distance correlations among zeros.
Furthermore, for finite $N$-cases, the repulsion between these
zeros for small distances is stronger for localized case than the
delocalized case. 
The presence of the repulsion leads zeros of the
delocalized case to
distribute uniformly in the phase space but the distribution is not a random
one. In the localized case too, the existence of a stronger repulsion
 may be the cause of the pattern formation of zeros.  
Again the presence of a symmetry affects the spacing distribution 
 for small-$s$ values, by increasing the probability of small-spacings 
relative to no-symmetry case (figure 6(d)); this happens due to
many very close-lying 
zeros on the symmetry-axis. But our analysis for various 
$N$-values indicates the tendency of this probability to decrease with 
higher $N$, approaching the same behaviour as in no-symmetry case 
(i.e eq.(14)) in $N\rightarrow\infty$ limit. This is what we expect
too as the fraction 
of zeros ($N^(-1/2)$) on the symmetry-axis  vanishes in the semiclassical 
regime. The fitting analysis followed by $\chi^2$-analysis further confirms 
that, in $N\rightarrow\infty$-limit, $P(s)$ is not affected by the presence 
or absence of a symmetry, localization or delocalization of quantum
dynamics and can be described by eq.(14). (The calculated
$\chi^2$-value, if eq.(14) is taken as hypothesis, is $12.56$,
$18.24$ and $30.78$ for 14 degrees of freedom of
1-symmetry, 2-symmetry and localized cases respectively. 
 This is much less than the theoretical $\chi^2$ $(=23.68)$ at $5\%$level
of significance for the first two cases, thus indicating
correctness of our hypothesis. For the last case although the observed
value is much greater than the theoretical one even for $N=199$
but again the decreasing $\chi^2$ with increasing $N$ (e.g
$\chi^2=199.65$ for $N=99$ and $14$ degrees of freedom) suggests
the validity of eq.(14) in $N\rightarrow\infty$ limit for this
case too.\par 

The $R_2(r)$-study does not inform us much about the effect of various
parameters on long range correlations. We study, therefore, another 
  fluctuation measure, namely, the "number variance"
$n(r)$ which is defined as the variation in the number
of zeros in an area of size $r^2$, where $r^2$ is sufficiently large to
include many zeros on an average and is 
two-dimensional analogue of the number variance quite often used for
spectrum studies. 
For a random distribution of
points distributed on a unit plane (with no correlation among
them), the variance increases as $r^2$. 
If the variance remains smaller than $r^2$ as the mean number of
zeros increases, the existence of long-range correlation in the
spectrum is indicated. 
The figure 7. shows that the variance in the
number of zeros for large $r$ is increasing with $r$; the 
rate of increase, for both QKR and RF,
is much slower than  that of the random distribution of points. 
This reflects that the
longe range correlations are indeed present among zeros of both
QKR and RF, although stronger  
in the former as compared to latter, if the dynamics is
localized or symmetries are present (figures 7(a),(b),(c)); 
the reverse is true if the
dynamics is delocalized with no symmetry (figure 7(a)). 
As can be seen from figure 7(d), these correlations
are strongest for delocalized dynamics with no symmetry, a little
weaker for symmetry cases and weakest
for localized case. A similar tendency was seen in case of
the eigenvalues too [5,12]; the presence of a symmetry or localization of
dynamics results in weakening of the long range correlations with a
tendency to behave as in the case of Poission distribution. (This
should not be surprising as Poission distribution results due to a
large number of symmetries present in the system). \par 
	In this paper, we have attempted to gain an insight in the
	complex world of wavefuctions by numerically analyzing the
	 distribution 
of their zeros. Although the studies here have been carried out for 
one system only, we believe in general applicability of results
obtained  here. Our study indicates that the zeros for 
strongly chaotic cases tend to accquire uniform distribution in the phase
space in the limit $N\rightarrow\infty$; this distribution is
different from that of the set of random points. The analysis of
various fluctuation
	measures shows that, in the large $N$-limit, the short-range 
correlations of zeros of quantum chaotic systems are universal,  
parametric-independent and have the same nature 
as those for random functions case or random polynomials but long-range 
correlations seem to differ. We expect the nature of this deviation to 
be system-dependent; this expectation has its root in the earlier observation 
of similar tendency in the eigenvalues-case.  
	Again, the localization vs
	delocalization phenomena of quantum state reveals itself in 
	the statistical
	behaviour of zeros too. The presence of a symmetry induces the 
	changes in the distribution of zeros and also modifies the 
long range correlations. This again confirms that zeros contain
	a lot of information about the underlying classical as well as
	quantum dynamics and  therefore can be
	used for a better understanding of the subject. But we have not 
	as yet extracted all the information about zeros.
	For example, we still have to understand as to how the 
motion of zeros is affected by the variation of various parameters of
the system. We also need to analyze higher order correlation which
can tell us more about their parametric and system dependence. 
 It is also desirable to have a complete 
analytical formulation of statistical properties of these zeros as it 
will help us in a better understanding of wavefunctions.\par

\bigskip
\noindent {\bf Acknowledgements}

	I am grateful to A.Voros for suggesting
the problem and for the
numerical program to compute the zeros in phase space. 
I also wish to express my gratitude to P.Leboeuf,
and O.Bohigas for many fruitful discussions and Jawaharlal Nehru Centre 
for advanced research (JNCASR) for their Computing facility
and economic support.

\vskip 5mm
\centerline{{\bf References}}
\vglue 2truecm

\item{$\lbrack$1$\rbrack$} M.V.Berry, J. Phys. A10, 2083 (1977).\par
\item{$\lbrack$2$\rbrack$} E.J. Heller, Phys. Rev. Lett. 53, 1515,
(1984); also "in Quantum Chaos and Statistical Nuclear Physics, 
proccuernavaca, Mexico, 1986, (eds. T.H.Seligman and H.Hishioka, Lecture
notes in Physics 263, Springer: Berlin).\par
\item{$\lbrack$3$\rbrack$} 
E.B.Bogomolny, Physica D, 31, 169 (1988).
\item{$\lbrack$4$\rbrack$}  
G.Casati and L.Molinari, Prog.Theor.Phys.Suppl.,98 287 (1989).
\item{$\lbrack$5$\rbrack$} 
P.Shukla, Ph.D Thesis, (Jawahar Lal Nehru University, New Delhi,
1992).
\item{$\lbrack$6$\rbrack$} 
F.Haake, "Quantum Signature of Chaos", (Springer, Berlin 1991).
\item{$\lbrack$7$\rbrack$} 
P.Leboeuf and A.Voros, J.Phys. A 23, (1990); "Quantum Chaos", 
(editiors: G.Casati and B.Chirikov, Cambridge University Press), Cambridge.
\item{$\lbrack$8$\rbrack$} 
P.Leboeuf, J.Phys.A 24, 4575, (1991).
\item{$\lbrack$9$\rbrack$} 
E.Bogomolny, O.Bohigas and P.Leboeuf, Phys. Rev. Lett., 68 2726,
(1992).
\item{$\lbrack$10$\rbrack$} 
P.Leboeuf and P.Shukla, J.Phys.A. 29 (1996).
\item{$\lbrack$11$\rbrack$} 
F.M.Izrailev, Phys. Rev. Lett. 56, 541 (1986).
\item{$\lbrack$12$\rbrack$} 
A.Pandey, R.Ramaswamy and P.Shukla, Pramana (Indian J.Phys.), 41 L75
(1993);
P.Shukla and A.Pandey, to appear in Nonlinearity (IOP, U.K.).
\item{$\lbrack$13$\rbrack$} 
A.Voros, Phys. Rev. A40, 6814, (1989).
\item{$\lbrack$14$\rbrack$} W.H.Press et.al., "Numerical Recipes",
Cambridge University Press, Cambridge.

\vskip 5mm
\centerline{{\bf Figure Captions}}
\vglue 2truecm

\noindent{\bf Figure 1.}: The superposition of the zeros of all
eigenfunctions of $U$, \par
(a) $K=20000, M=1, N=99, \gamma =0.7071, \theta _0 =\pi/2N $,\par
(b) $K=10, M=30, N=99, \gamma =0.7071, \theta _0 =\pi/2N $.\par
\bigskip
\noindent{\bf Figure 2.}: The Husimi distribution of a single eigenfunction,
\par
(a) $K=10, M=30, N=61, \gamma =0.7071, \theta _0 =\pi/2N $,\par
(b) $K=10, M=30, N=61, \gamma =0.0, \theta _0 =0.0 $,\par
(c) $K=10, M=30, N=61, \gamma =0.0, \theta _0 =0.0 $,\par
(d) $K=20000, M=1, N=61, \gamma =0.7071, \theta _0 =\pi/2N $.\par
\bigskip
\noindent{\bf Figure 3.}: The density $\rho(q,p)$ of zeros in the phase
plane with respect to q (averaged over all values of p and measured in units
of $<q>$) and p (averaged over all values of q and measured in units
of $<p>$),\par
(a) $\&$ (b) 
$K=20000, M=1, N=99, \gamma =0.7071, \theta _0 =\pi/2N $, (solid line),
$K=10, M=30, N=99, \gamma =0.7071, \theta _0 =\pi/2N $, (dashed line),
 RFC case  (dotted line),\par
(c) $\&$ (d) 
$K=20000, M=1, N=99, \gamma =0.0, \theta _0 =\pi/2N $, (solid
line), RFR case (dotted line),\par

\bigskip
\noindent{\bf Figure 4.}: The superposition of the zeros of all
eigenfunctions of $U$,\par 
(a) $K=10, M=30, N=61, \gamma =0.0, \theta _0 =0.0 $,\par
(b) $K=10, M=30, N=61, \gamma =0.0, \theta _0 =\pi/2N $.\par
\bigskip
\noindent{\bf Figure 5.} The histogram for Two-point correlation 
function $R_2$ of the zeros (with unit mean spacing),\par 
(a) Delocalized without any symmetry (solid histogram) and 
RFC (dotted histogram),\par
(b) Delocalized with one symmetry (solid histogram) and RFR 
(dotted histogram),\par
(c) Delocalized with two symmetries  (solid histogram) and RFA 
(dotted histogram),\par
(d) with respect to small values of $r$, no symmetry case (solid
line), 1-symmetry (dot line), 2-symmetry (small-dash line), localized
case (big-dash line). Note here the difference between various
curves is bigger than the finite-smaple error ($\approx 5\times
10^{-3}$) associated with each case.\par
\bigskip
\noindent{\bf Figure 6.} The Histogram for Nearest-Neighbour
spacing distribution, (with unit mean spacing)\par
(a) Delocalized without any symmetry (solid histogram) and 
RFC (dotted histogram),\par
(b) Delocalized with one symmetry (solid histogram) and RFR 
(dotted dotted histogram),\par
(c) Delocalized with two symmetries (solid histogram) and RFA 
(dotted histogram),\par
(d) Localized case with no symmetry (solid histogram) and Delocalized
case with no symmetry (dotted histogram),\par 
Here solid curve in (a),(b),(c) is the fitting given by eq.(14). \par
\bigskip
\noindent{\bf Figure 7.} Number Variance $n(r)$,\par
(a) Delocalized without any symmetry (solid curve), Localized case
(dashed curve) and RFC (dotted curve),\par
(b) Delocalized with one symmetry (solid curve) and RFR 
(dotted curve),\par
(c) Delocalized with two symmetries (solid curve) and RFA 
(dotted curve).\par

\bye